\documentclass[floats, eqnum, showpacs, nofootinbib, 
,11pt
,preprint 
,eqsecnum 
]{revtex4-1}

\usepackage{color,graphicx}
\usepackage{amsfonts}
\usepackage{amssymb}
\begin{document}

\title{
A No-Go Theorem for Rotating Stars of a Perfect Fluid without Radial Motion in Projectable Ho\v{r}ava--Lifshitz Gravity}

\author{Naoki Tsukamoto
\footnote{Electronic
address:11ra001t@rikkyo.ac.jp}
and Tomohiro Harada
}

\affiliation{
Department of Physics, Rikkyo University, 3-34-1 Nishi-Ikebukuro, Toshima-ku, Tokyo~171-8501,~Japan
}
\date{\today}

\begin{abstract}
Ho\v{r}ava--Lifshitz gravity has covariance only under the foliation-preserving diffeomorphism.
This implies that the quantities on the constant-time hypersurfaces should be regular.
In the original theory, the projectability condition, which strongly restricts the lapse function, is proposed.
We assume that a star is filled with a perfect fluid with no-radial motion 
and that it has reflection symmetry about the equatorial plane.
As a result, we find 
a no-go theorem for stationary and axisymmetric star solutions in projectable Ho\v{r}ava--Lifshitz gravity 
under the physically reasonable assumptions in the matter sector.
Since we do not use the gravitational action to prove it,
our result also works out in other projectable theories
and applies to not only strong gravitational fields, but also weak gravitational ones. 
\end{abstract}


\preprint{RUP-11-5}

\maketitle


\section{Introduction}

Recently, Ho\v{r}ava proposed a power-counting renormalizable gravitational theory \cite{horava2009_lifshitz,horava2009_membranes}.
The theory is called Ho\v{r}ava--Lifshitz gravity, because it exhibits the Lifshitz-type anisotropic scaling in the ultraviolet:
\begin{eqnarray}
t \rightarrow b^zt ,\quad x^i\rightarrow bx^i
\end{eqnarray}
where $t$, $x^{i}$, $b$ and $z$ are the temporal coordinate, the spatial coordinates, the scaling factor and the dynamical critical exponent, respectively,
and $i$ runs over one, two and three.
Since this theory is expected to be renormalizable and unitary, 
its phenomenological aspects \cite{weinfurtner_sotiriou_visser2010,gumrukcuoglu_mukohyama2011} 
and variants \cite{sotiriou_visser_weinfurtner2009lett,sotiriou2011}
strenuously have been investigated, including 
black holes \cite{harada_miyamoto_tsukamoto2011,park2009,lu_mei_pope2009,
tang_chen_2010,cai_cao_ohta2009,cai_cao_ohta2009B,Barausse_Sotiriou_2013,Wang_2013,Barausse_Sotiriou_2012,Wang_2012},
dark matter \cite{mukohyama2009,mukohyama2010}, 
dark energy \cite{saridakis2010},
the solar system test~\cite{harko_kovacs_lobo2011} 
and so on.

The field variables in this theory are the lapse function, $N(t)$, the shift vector, $N^{i}(t,x)$, and the spatial metric, $g_{ij}(t,x)$.
Note that the shift vector, $N^{i}$, and the spatial metric, $g_{ij}$, can depend on both $t$ and $x^{i}$, but that the lapse function, $N$, can only do so on $t$.
Since the lapse function, $N$, can be interpreted as a gauge field associated with the time reparametrization, it is natural to restrict it to be space independent.
This assumption, called the projectability condition, is proposed in Ho\v{r}ava's original paper \cite{horava2009_lifshitz} 
from the view point of quantization.
However, the pathological behaviors of the projectability condition, such as the infrared instability and the strong coupling, 
are found \cite{Horava_2009_Lett,horava2009_lifshitz,horava2009_membranes,Wang_Maartens_2009,Koyama_Arroja_2009,Wang_Wu_2011,Charmousis_Niz_Padilla_Saffin_2009,Blas_Pujolas_Sibiryakov_2009},
and the theory has been extended to avoid the adverse situation~\cite{Blas_Pujolas_Sibiryakov_2010,Horava_Melby-Thompson_2010}.

Since higher derivative terms do not contribute at large distances,
the action of this theory can recover the apparent form of general relativity
if we tune a coupling parameter.
In this context, it seems that projectable Ho\v{r}ava--Lifshitz gravity passes astrophysical tests.
However, we will show that, actually, this is not true in this paper.

In this theory, black holes have been investigated eagerly, 
while stars have not been studied so much \cite{izumi_mukohyama2010,greenwald_papazoglou_wang2010}.
The comparison of the features of star solutions in Ho\v{r}ava--Lifshitz gravity with the corresponding ones in Einstein gravity 
would be one of the astrophysical tests for Ho\v{r}ava--Lifshitz gravity.
It is important to investigate star solutions, gravitational collapse \cite{Greenwald_Lenells_Satheeshkumar_Wang_2013} 
and the formation of black holes.

The first study of stars in Ho\v{r}ava--Lifshitz gravity was done by Izumi and Mukohyama \cite{izumi_mukohyama2010}.
They found a no-go theorem that 
no spherically symmetric and static solution filled a perfect fluid without radial motion exists in this projectable theory 
under the assumptions that 
the energy density is a piecewise-continuous and non-negative function of the pressure and that the pressure at the center is positive.
Their result is powerful, because it does not depend on the gravitational action.

To construct star solutions, we have to change at least one of their assumptions 
for the matter sector, the symmetry of spacetime, the projectability and the invariance under the foliation-preserving diffeomorphism. 
Greenwald, Papazoglou and Wang found spherically symmetric static solutions, 
which are filled with a perfect fluid with radial motion and a class of an anisotropic fluid in the projectable Ho\v{r}ava--Lifshitz gravity 
without the detailed balance condition \cite{greenwald_papazoglou_wang2010}.

It seems that static solutions are too simple to describe realistic stars, which are generally rotational.
In this paper, we investigate a stationary and axisymmetric star in projectable Ho\v{r}ava--Lifshitz gravity.
We find a no-go theorem that the stationary and axisymmetric star filled with a perfect fluid without radial motion
in the reflection symmetry about the equatorial plane 
does not exist 
under the physically reasonable conditions on the matter sector.
Since we do not use the gravitational action to prove it,
our result also works out in other projectable theories \cite{sotiriou_visser_weinfurtner2009lett,sotiriou_visser_weinfurtner2009jhep}
and applies to not only strong gravitational fields, like neutron stars, but also weak gravitational ones, like planets or moons. 
Our proof implies another ill behavior of the projectability condition 
if we follow a principle that stars should be described by stationary solutions of a low-energy effective theory. 
On the other hand, even if we do not follow this principle, our result would be useful to investigate rotating-star solutions in this theory and
then to compare the solutions with the corresponding ones in Einstein gravity for astrophysical tests of this~theory.
 
This paper is organized as follows.
In Section 2, we shall describe the definitions, the basic equations and the properties of Ho\v{r}ava--Lifshitz gravity.
In Section 3, we give the main result that there are no stationary and axisymmetric star solutions 
with a perfect fluid, which does not have the radial component of the four-velocity
under a set of reasonable assumptions in the matter sector.
In Section 4, we summarize and discuss our result.
In Appendix A, we show the explicit expression for the equation of motion.
In Appendix B, we show the triad components of the extrinsic curvature tensor.
In this paper, we use the units in which $c=1$.


\section{Properties of Ho\v{r}ava--Lifshitz Gravity}

In this section, we shall describe the definitions, the basic equations and the properties of Ho\v{r}ava--Lifshitz gravity.
Ho\v{r}ava--Lifshitz gravity does not have general covariance, 
since the Lifshitz-type anisotropic scaling treats time and space differently.
Instead, this theory is invariant under the foliation-preserving diffeomorphism:
\begin{eqnarray}
\; t\rightarrow \tilde{t}(t),\; x^{i}\rightarrow \tilde{x}^{i}(t,x^{j})
\end{eqnarray}
This means that the foliation of the spacetime manifold by constant-time hypersurfaces has a physical meaning.
Thus, the quantities on the constant-time hypersurfaces, such as the extrinsic curvature tensor and the shift vector, must be regular.

It is useful to describe the line element in the Arnowitt--Deser--Misner (ADM) form \cite{misner_thorne_wheeler1973}:
\begin{equation}
 ds^2=-N^2dt^2+g_{ij}(dx^i+N^idt)(dx^j+N^jdt)
\end{equation}
The action proposed by Ho\v{r}ava \cite{horava2009_lifshitz}
is given by:
\begin{eqnarray}
I=I_{g}+I_{m}
\end{eqnarray}
\begin{eqnarray}\label{eq:gravitational_action}
I_{g}
&=&\int dtd^3x\sqrt{g}N \left\{ \frac{2}{\kappa^2}(K^{ij}K_{ij}-\lambda K^2)-\frac{\kappa^2}{2\omega^4}C_{ij}C^{ij} 
+\frac{\kappa^2\mu}{2\omega^2} \varepsilon^{ijk}R_{il} D_jR^l_k -\frac{\kappa^2\mu^2}{8}R_{ij}R^{ij} \right. \nonumber\\
&& \left. +\frac{\kappa^2\mu^2}{8(1-3\lambda)}\left(\frac{1-4\lambda}{4}R^2+ \Lambda_W R -3\Lambda^2_W \right) \right\}
\end{eqnarray}
where $I_{m}$ is the matter action, $R$ is the Ricci scalar of $g_{ij}$, $R_{ij}$ is the Ricci tensor of $g_{ij}$,
$D_{i}$ is the covariant derivative compatible with $g_{ij}$, $K_{ij}$ is the extrinsic curvature of a constant-time hypersurface, defined by:
\begin{eqnarray}\label{eq:extrinsic-curvature}
K_{ij}=\frac{1}{2N}(\partial_{t}{g}_{ij}-D_{i}N_{j}- D_{j}N_{i} )
\end{eqnarray}
$K=g^{ij}K_{ij}$, $C_{ij}$ is the Cotton tensor, defined by:
\begin{eqnarray}
C^{ij}=\varepsilon^{ikl}D_k\left(R^j_l -\frac{1}{4}R\delta^j_l \right)
\end{eqnarray}
$\varepsilon^{ikl}=\epsilon^{ikl}/\sqrt{g}$ is the antisymmetric tensor, which is covariant with respect to $g_{ij}$,
and $\kappa, \omega, \mu, \lambda$ and $\Lambda_W$ are constant parameters.
We can rewrite the gravitational action (\ref{eq:gravitational_action}):
\begin{eqnarray}
I_{g}
=\int dtd^3x\sqrt{g}N[\alpha (K^{ij}K_{ij}-\lambda K^2)+\beta C_{ij}C^{ij} 
+\gamma \varepsilon^{ijk}R_{il} D_jR^l_k+\zeta R_{ij}R^{ij}+\eta R^2+\xi R+\sigma]
\end{eqnarray}
where parameters $\alpha, \beta, \gamma, \zeta, \eta, \xi$ and $\sigma$ are given by:
\begin{eqnarray}
\alpha=\frac{2}{\kappa^2},\quad \beta=-\frac{\kappa^2}{2\omega^4},\quad \gamma=\frac{\kappa^2\mu}{2\omega^2},\quad \zeta=-\frac{\kappa^2\mu^2}{8},\nonumber\\ 
\eta=\frac{\kappa^2\mu^2}{8(1-3\lambda)}\frac{1-4\lambda}{4},\quad \xi=\frac{\kappa^2\mu^2}{8(1-3\lambda)}\Lambda_W, \quad \sigma=\frac{\kappa^2\mu^2}{8(1-3\lambda)}(-3\Lambda^2_W)
\end{eqnarray}
If we take $\lambda =1$ to recover the apparent form of general relativity and the apparent Lorentz invariance,
we can compare this action to that of general relativity.
Then, we obtain:
\begin{eqnarray}
\alpha=\frac{1}{16\pi G},\quad \xi=\alpha,\quad \sigma=-2\Lambda \alpha
\end{eqnarray}
where $\Lambda$ is the cosmological constant and $G$ is Newton's constant.

Under the infinitesimal coordinate transformation:
\begin{eqnarray}
\delta t =f(t),\qquad \delta x^{i}=\zeta^{i}(t,x)
\label{eq:foliation-preserving-diffeomorphism}
\end{eqnarray}
$g_{ij}$, $N^{i}$ and $N$ transform as:
\begin{eqnarray}
&&\delta g_{ij}=f\partial_{t}g_{ij}+ \mathcal{L}_{\zeta}g_{ij} \\ 
&&\delta N^{i}=\partial_{t}(N^{i}f)+\partial_{t}\zeta^{i}+\mathcal{L}_{\zeta}N^{i} \\
&&\delta N_{i}=\partial_{t}(N_{i}f)+g_{ij}\partial_{t}\zeta^{j}+\mathcal{L}_{\zeta}N_{i} \\
&&\delta N=\partial_{t}(Nf) 
\end{eqnarray}
where $\mathcal{L}_{\zeta}$ is the Lie derivative along $\zeta^{i}(t,x)$.
$\mathcal{L}_{\zeta}g_{ij}$ and $\mathcal{L}_{\zeta}N^{i}$ are given by:
\begin{eqnarray}
\mathcal{L}_{\zeta}g_{ij}=g_{jk}D_{i}\zeta^{k}+g_{ik}D_{j}\zeta^{k} \\
\mathcal{L}_{\zeta}N^{i}=\zeta^{k}D_{k}N^{i}-N^{k}D_{k}\zeta^{i}
\end{eqnarray}

By the variation of the action with respect to $N$, we get the Hamiltonian constraint:
\begin{eqnarray}\label{eq:Hamiltonian-constraint}
H_{g\bot }+H_{m\bot }
=0
\end{eqnarray}
where:
\begin{eqnarray}
H_{g\bot }&\equiv& -\frac{\delta I_{g}}{\delta N} \nonumber\\
&=&\int dx^3\sqrt{g} \left[(\alpha K^{ij}K_{ij}-\lambda K^{2})-\beta C_{ij}C^{ij} 
-\gamma \varepsilon^{ijk}R_{il}D_{j}R^{l}_{k}-\zeta R_{ij}R^{ij}-\eta R^{2}-\xi R-\sigma \right]\qquad 
\end{eqnarray}
and:
\begin{eqnarray}
H_{m\bot }\equiv -\frac{\delta I_{m}}{\delta N}
=\int dx^{3}\sqrt{g}T_{\mu\nu}n^{\mu}n^{\nu}
\end{eqnarray}
Here, $n^{\mu}$ is defined as:
\begin{eqnarray}
&&n_{\mu}dx^{\mu}=-Ndt ,\quad n^{\mu}\partial_{\mu}=\frac{1}{N}(\partial_{t}-N^{i}\partial_{i})
\end{eqnarray}
Notice that due to the projectability condition $N=N(t)$, the Hamiltonian constraint is global in \linebreak Ho\v{r}ava--Lifshitz gravity,
while it is local in general relativity.

From the variation of the action with respect to $N^{i}$, we obtain the momentum constraint:
\begin{eqnarray}\label{eq:momentum-constraint}
\mathcal{H}_{gi}+\mathcal{H}_{mi}
=0 
\end{eqnarray}
where:
\begin{eqnarray}
&&\mathcal{H}_{gi}\equiv -\frac{1}{\sqrt{g}}\frac{\delta I_{g}}{\delta N^{i}}=-2\alpha D^{j}(K_{ij}-\lambda Kg_{ij}) \\
&&\mathcal{H}_{mi}\equiv -\frac{1}{\sqrt{g}}\frac{\delta I_{m}}{\delta N^{i}}=T_{i\mu}n^{\mu}
\end{eqnarray}

By the variation of the action with respect to $g_{ij}$, we get the equation of motion:
\begin{eqnarray}\label{eq:equation-of-motion}
\mathcal{E}_{gij}+\mathcal{E}_{mij}
=0
\end{eqnarray}
where:
\begin{eqnarray}
\mathcal{E}_{gij}
\equiv g_{ik}g_{jl}\frac{2}{N\sqrt{g}}\frac{\delta I_{g}}{\delta g_{kl}} \\
\mathcal{E}_{mij}\equiv g_{ik}g_{jl}\frac{2}{N\sqrt{g}}\frac{\delta I_{m}}{\delta g_{kl}}
=T_{ij}
\end{eqnarray}
The explicit expression for the equation of motion is given in Appendix A.

By the invariance of the gravitational action and the matter action under the infinitesimal transformation (\ref{eq:foliation-preserving-diffeomorphism}),
we get the energy conservation:
\begin{eqnarray}\label{eq:energy-conservation}
N\partial_{t}H_{\alpha \bot}+\int {dx^{3} \left( N^{i}\partial_{t}(\sqrt{g}\mathcal{H}_{\alpha i})+\frac{N\sqrt{g}}{2}\mathcal{E}_{\alpha}^{\; ij}\partial_{t}g_{ij} \right) }
=0 \nonumber\\
\end{eqnarray}
and the momentum conservation:
\begin{eqnarray}\label{eq:momentum-conservation}
0
=\frac{1}{N}(\partial_t-N^jD_j)\mathcal{H}_{\alpha i}+K\mathcal{H}_{\alpha i} 
-\frac{1}{N}\mathcal{H}_{\alpha j}D_iN^j-D^j\mathcal{E}_{\alpha ij}
\end{eqnarray}
where $\alpha$ represents $g$ or $m$.

In the next section, we will only use the momentum conservation of the matter to show that no stationary and axisymmetric star solution exists.
Therefore, our result does not depend on the gravitational~action.


\section{No Stationary and Axisymmetric Star Solutions}

In this section, 
we show a no-go theorem for stationary and axisymmetric star solutions in projectable Ho\v{r}ava--Lifshitz gravity.
To prove it, we assume that a star is filled with a perfect fluid, which does not have the radial component of the four-velocity,
that it has the reflection symmetry about the equatorial plane,
that the energy density is a piecewise-continuous and non-negative function of the pressure,
that the pressure is a continuous function of $r$
and that the pressure at the center of the star is positive.


\subsection{Stationary and Axisymmetric Configuration}

We consider stationary and axisymmetric configurations with the timelike and spacelike Killing vectors, respectively, given by:
\begin{eqnarray}
&&t^{\mu}\partial_{\mu}=\partial_{t} \\\label{eq:spacelike_killing_vector}
&&\phi^{\mu}\partial_{\mu}=\partial_{\phi}
\end{eqnarray}

Under the stationary configurations, the lapse function, $N$, does not depend on $t$.
In the original theory, the projectability condition $N=N(t)$ is proposed \cite{horava2009_lifshitz}.
This condition means that the lapse function, $N$, does not depend on the spatial coordinates, $x^{i}$, 
but only can do so on the temporal coordinate, $t$.
Thus, the lapse function, $N$, is a constant.

The timelike Killing vector, $t^{\mu}$, implies everywhere:
\begin{eqnarray}
N^{2}-N_{i}N^{i}>0
\end{eqnarray}
The spacelike Killing vector, $\phi^{\mu}$, implies that:
\begin{eqnarray}
\phi^{\mu}\phi_{\mu}=g_{\phi\phi}
\end{eqnarray}
is a geometrical invariant.

As a part of the gauge condition, we take:
\begin{eqnarray}
g_{r\theta}=g_{r\phi}=0
\end{eqnarray}
Under this gauge condition, the general form for the spatial line element is described by \cite{bardeen_piran1983}:
\begin{eqnarray}\label{eq:spatial-line-element}
dl^{2}=\psi^{4}[A^{2}dr^{2}+\frac{r^{2}}{B^{2}}d\theta^{2}+r^{2}B^{2}(\sin\theta d\phi+\xi d\theta)^{2}]
\end{eqnarray}
where $\psi$, $A$, $B$ and $\xi$ are functions of $r$ and $\theta$, but neither $t$ nor $\phi$ for stationarity and axisymmetry.

Now we assume that the spacetime has a rotation axis, where $\sin\theta=0$. 
This means:
\begin{eqnarray}\label{eq:rotation-axis}
\phi^{\mu}\phi_{\mu}=g_{\phi\phi}=0
\end{eqnarray}
there \cite{hayward2000}.


\subsection{Triad Components of Shift Vector}

We define triad basis vectors \textrm{\boldmath $\left\{ e_{(i)} \right\}$}. 
\textrm{\boldmath $e_{(1)}$} is along the radial direction; \textrm{\boldmath $e_{(3)}$} is along the axial Killing vector 
and \textrm{\boldmath $e_{(2)}$} is fixed by the orthonormality and the right-hand rule.
The coordinate components for the orthonormal triad are:
\begin{eqnarray}\label{eq:triad-one}
&&e_{(1)}^{i}=\frac{1}{\psi^{2}} \left[ \frac{1}{A},0,0 \right] \\\label{eq:triad-two} 
&&e_{(2)}^{i}=\frac{1}{\psi^{2}} \left[ 0,\frac{B}{r},-\frac{\xi B}{r\sin\theta} \right] \\\label{eq:triad-three}
&&e_{(3)}^{i}=\frac{1}{\psi^{2}} \left[ 0,0,\frac{1}{rB\sin\theta} \right]
\end{eqnarray}
where we have used the spatial line element (\ref{eq:spatial-line-element}).
The projection of the shift vector on the triad is related to its coordinate components by:
\begin{eqnarray}\label{eq:shift-one}
&&N_{(1)}=\frac{N_{r}}{\psi^{2}A}\\\label{eq:shift-two}
&&N_{(2)}=\frac{N_{\theta}B}{\psi^{2}r}-\frac{N_{\phi}\xi B}{\psi^{2}r\sin\theta}\\\label{eq:shift-three}
&&N_{(3)}=\frac{N_{\phi}}{\psi^{2}rB\sin\theta}
\end{eqnarray}


\subsection{Regularity Conditions at the Origin}

Here, we give the regularity conditions of the shift vector, $N^{i}$, near the origin.
A tensorial quantity is regular at $r=0$ if and only if 
all its components can be expanded in non-negative integer powers of $x$, $y$ and $z$ in locally Cartesian coordinates, defined by:
\begin{eqnarray}
&&x\equiv r\sin\theta\cos\phi \\
&&y\equiv r\sin\theta\sin\phi \\
&&z\equiv r\cos\theta 
\end{eqnarray}
The Lie derivative of the shift vector, $N^{i}$, along the spacelike Killing vector vanishes, or:
\begin{eqnarray}\label{eq:Lie-derivative-shift-vector}
N^{i}_{\;,j}\phi^{j}-\phi^{i}_{\; ,j}N^{j}=0
\end{eqnarray}
In locally Cartesian coordinates, the spacelike Killing vector is written as:
\begin{eqnarray}
\phi^{i}\partial_{i}=-y\partial_{x}+x\partial_{y}
\end{eqnarray}
Then, its components of Equation (\ref{eq:Lie-derivative-shift-vector}) are:
\begin{eqnarray}
&&-N^{x}_{\;,x}y+N^{x}_{\;,y}x+N^{y}=0 \\
&&-N^{y}_{\;,x}y+N^{y}_{\;,y}x-N^{x}=0 \\
&&-N^{z}_{\;,x}y+N^{z}_{\;,y}x=0 
\end{eqnarray}
The general regular solution of these equations is:
\begin{eqnarray}
&&N^{x}=F_{1}(z,\rho^{2})x-F_{2}(z,\rho^{2})y \\
&&N^{y}=F_{1}(z,\rho^{2})y+F_{2}(z,\rho^{2})x \\
&&N^{z}=F_{3}(z,\rho^{2}) 
\end{eqnarray}
where $F_{1}$, $F_{2}$ and $F_{3}$ are independent and regular functions, which depend on $z$ and $\rho^{2} \equiv x^{2}+y^{2}$.

Now, transforming $N^{i}$ back to the spherical coordinates, $r,\theta$ and $\phi$, we get the spherical components:
\begin{eqnarray}\label{eq:r-components-shift-vector}
&&\frac{N^{r}}{r}=\sin^{2}\theta F_{1}+\frac{1}{r}\cos\theta F_{3} \\\label{eq:theta-components-shift-vector}
&&\frac{N^{\theta}}{\sin\theta}=\cos\theta F_{1}-\frac{F_{3}}{r} \\\label{eq:phi-components-shift-vector}
&&N^{\phi}=F_{2} 
\end{eqnarray}
On the rotation axis ($\sin\theta=0$), thus, we obtain: 
\begin{eqnarray}\label{eq:N_theta}
N^{\theta}=0
\end{eqnarray}
Using Equations (\ref{eq:spatial-line-element}), (\ref{eq:triad-one})--(\ref{eq:triad-three}) and (\ref{eq:r-components-shift-vector})--(\ref{eq:phi-components-shift-vector}), 
the triad components are given by:
\begin{eqnarray}
&&N_{(1)}=\psi^{2}A(r\sin^{2}\theta F_{1}+\cos \theta F_{3}) \\
&&N_{(2)}=\frac{\psi^{2}}{B}\sin\theta (r\cos \theta F_{1}-F_{3}) \\
&&N_{(3)}=\psi^{2}B\sin\theta(r\xi \cos\theta F_{1}-\xi F_{3}+rF_{2})
\end{eqnarray}
Here, we additionally assume the reflection symmetry about the equatorial plane $z=0$ or $\theta=\pi /2$.
Then, $N^{x}$ and $N^{y}$ must be even functions of $z$, and $N^{z}$ must be an odd function of $z$.
This implies that $F_{1},F_{2}$ must be even functions of $z$, and $F_{3}$ must be an odd function of $z$.
Since $N^{r}$ is an odd function of $z$ on the rotation axis ($\sin\theta=0$), we get: 
\begin{eqnarray}\label{eq:N_r}
N^{r}=0
\end{eqnarray}
at the origin.


\subsection{Matter Sector and Momentum Conservation}

For simplicity, we assume that the matter consists of a perfect fluid.
The stress-energy tensor is given~by:
\begin{eqnarray}
T_{\mu\nu}=(\rho+P)u_{\mu}u_{\nu}+Pg_{\mu\nu}
\end{eqnarray}
where $P$ and $\rho$ represent the pressure and the energy density, respectively.
We assume the four-velocity given by:
\begin{eqnarray}
u^{\mu}\partial_{\mu}
&=&\frac{1}{D}(t^{\mu}+\omega \phi^{\mu})\partial_{\mu}\nonumber\\
&=&\frac{1}{D}\partial_{t}+\frac{\omega}{D}\partial_{\phi}
\end{eqnarray}
where:
\begin{eqnarray}
D\equiv (N^{2}-N_{i}N^{i}-2\omega N_{\phi}-\omega^{2}g_{\phi \phi})^{\frac{1}{2}}
\end{eqnarray}
is the normalization factor and $\omega$ is a function of $r$ and $\theta$.
For the four-velocity, $u^{\mu}$, to be timelike, we shall have $N^{2}-N_{i}N^{i}-2\omega N_{\phi}-\omega^{2}g_{\phi \phi}>0$.

We set $\alpha=m$, and then, the momentum conservation equation (\ref{eq:momentum-conservation}) of the matter becomes:
\begin{eqnarray}
0
=-\frac{1}{N}N^jD_j(T_{i\mu}n^\mu)+KT_{i\mu}n^\mu 
-\frac{1}{N}T_{j\mu}n^{\mu}D_iN^j-D^jT_{ij} 
\end{eqnarray}
After some calculation, we obtain the $r$ component:
\begin{eqnarray}\label{eq:momentum-conservation-r-component-nonprojectable}
0
=-P_{,r}+\frac{\rho+P}{D^{2}}\left\{ \frac{1}{2}(N_{i}N^{i})_{,r}+\omega N_{\phi ,r}+\frac{1}{2}\omega^{2}g_{\phi\phi,r} 
+\frac{N_{,r}}{N}N^{r}N_{r}+\frac{N_{,\theta}}{N}N^{\theta}N_{r} \right\}
\end{eqnarray}
Now, we use the projectability condition $N=N(t)$.
As we mentioned above, the projectability condition means 
that the lapse function, $N$, does not depend on the spatial coordinates, $x^{i}$, 
but only can do on the temporal coordinate, $t$.
Thus, the $r$ component of the momentum conservation equation (\ref{eq:momentum-conservation-r-component-nonprojectable}) becomes:
\begin{eqnarray}\label{eq:momentum-conservation-r-component}
0
=-P_{,r}+\frac{\rho+P}{D^{2}}\left\{ \frac{1}{2}(-N^{2}+N_{i}N^{i})_{,r} 
+\omega N_{\phi ,r}+\frac{1}{2}\omega^{2}g_{\phi\phi,r} \right\} 
\end{eqnarray}
We do not use the $\theta$ and $\phi$ components to prove that no stationary and axisymmetric star exists.

Here, we concentrate on the $r$ component of the momentum conservation of the matter on the rotation axis $\sin\theta=0$.
On the rotation axis, $g_{\phi\phi}$ and $g_{\phi\phi ,r}$ vanish from Equation (\ref{eq:rotation-axis}).
From Equation (\ref{eq:shift-three}), the regularity of the triad component of the shift vector, $N_{(3)}$, implies:
\begin{eqnarray}\label{eq:N_phi}
N_{\phi}=0
\end{eqnarray}
on the rotation axis.
Thus, $N_{\phi ,r}=0$.
Thus, the $r$ component of the momentum conservation equation~(\ref{eq:momentum-conservation-r-component}) on the rotation axis becomes:
\begin{eqnarray}\label{eq:momentum-conservation-r-component-axis}
0
=-P_{,r}-\frac{1}{2}\frac{(\rho+P)(N^{2}-N_{i}N^{i})_{,r}}{N^{2}-N_{i}N^{i}}
\label{eq:P_r}
\end{eqnarray}


\subsection{Contradiction of Momentum Conservation}

We assume that the star has the reflection symmetry about the equatorial plane $\theta=\frac{\pi}{2}$,
that the energy density, $\rho$, is a piecewise-continuous and non-negative function of the pressure, $P$,
that the pressure, $P$ is a continuous 
function of $r$
and that the pressure at the center of the star $P_{c}\equiv P(r=0)$ is positive.
Thus, $\rho+P$ is a piecewise-continuous function of $r$.
We have assumed that the energy density, $\rho$, is non-negative everywhere and that the pressure at the center, $P_{c}$, is positive;
hence, $\rho+P$ is positive at the center.
We define $r_{s}$ as the minimal value of $r$ for which at least one of 
$\left. (\rho+P) \right|_{r=r_{s}}$, $\lim_{r\rightarrow r_{s}-0}(\rho+P)$ and $\lim_{r\rightarrow r_{s}+0}(\rho+P)$ is nonpositive.

Dividing the momentum conservation equation (\ref{eq:momentum-conservation-r-component-axis}) by $\frac{1}{2}(\rho+P)$, we have:
\begin{eqnarray}\label{eq:momentum-conservation-r-component-axis-2}
\left\{ \log \left(N^{2}-N_{i}N^{i} \right) \right\}_{,r} 
=-2\frac{P_{,r}}{\rho+P}
\end{eqnarray}
Under the assumption that the energy density is a function of the pressure, $\rho=\rho(P)$,
integrating the momentum conservation equation (\ref{eq:momentum-conservation-r-component-axis-2}) over the interval $0\leq r< r_{s}$, we obtain: 
\begin{eqnarray}\label{eq:momentum-conservation-r-component-axis-3}
\left. \log \left(N^{2}-N_{i}N^{i} \right) \right|_{r=r_{s}} 
- \left. \log \left( N^{2}-N_{i}N^{i} \right) \right|_{r=0} 
=-2\int^{P_{s}}_{P_{c}}\frac{dP}{\rho+P}
\end{eqnarray}
where $P_{s}\equiv P(r=r_{s})$. 

The definition of $r_{s}$ implies that at least one of 
$\left. (\rho+P) \right|_{r=r_{s}}$, $\lim_{r\rightarrow r_{s}-0}(\rho+P)$ and $\lim_{r\rightarrow r_{s}+0}(\rho+P)$ is nonpositive.
Since we have assumed that $P(r)$ is a continuous function
and that $\rho$ is non-negative everywhere,
$P_{s}=\lim_{r\rightarrow r_{s}-0}P=\lim_{r\rightarrow r_{s}+0}P$ is non-positive.
Thus, we get: 
\begin{eqnarray}
P_{s}\leq 0<P_{c}
\end{eqnarray}
This implies that the right-hand side of Equation (\ref{eq:momentum-conservation-r-component-axis-3}) is positive.
However, the left-hand side of Equation~(\ref{eq:momentum-conservation-r-component-axis-3}) is nonpositive, 
since we have the projectability condition $N=N(t)$ and we obtain from Equations (\ref{eq:N_theta}), (\ref{eq:N_r}) and (\ref{eq:N_phi}): 
\begin{eqnarray}
\left. N_{i}N^{i} \right|_{r=0}
=0
\end{eqnarray}
at the center of the star.
This contradicts that the right-hand side of Equation (\ref{eq:momentum-conservation-r-component-axis-3}) is positive.


\section{Discussion and Conclusions}

Ho\v{r}ava--Lifshitz gravity is only covariant under the foliation-preserving diffeomorphism.
This means that the foliation of the spacetime manifold by the constant-time hypersurfaces has a physical meaning.
As a result, the regularity condition at the center of a star is more restrictive than the one in a theory that has general covariance.

Under the assumption that a star is filled with a perfect fluid 
that has no radial motion,
that it has reflection symmetry about the equatorial plane 
and that the matter sector obeys the physically 
reasonable conditions,
we have shown that the momentum conservation is incompatible with the projectability condition and 
the regularity condition at the center for stationary and axisymmetric configurations.
Since we have not used the gravitational action to prove it,
our result is also true in other projectable theories
\cite{sotiriou_visser_weinfurtner2009lett,sotiriou_visser_weinfurtner2009jhep}.
Note that our result is true under not only strong-gravity circumstances, like neutron stars, but also weak-gravity ones, like planets or moons. 
However, it is not certain that star solutions can exist in non-projectable theories.
Since we have used the covariance under the foliation-preserving diffeomorphism, the projectability condition and the assumptions of the matter sector
to prove the no-go theorem for stationary and axisymmetric stars,
our proof will not apply 
if we do not assume all the above.

Izumi and Mukohyama found that no spherically symmetric and static solution 
filled with a perfect fluid without radial motion 
exists in this theory 
under the assumption that the energy density is a piecewise-continuous and non-negative function of the pressure
and that the pressure at the center is positive \cite{izumi_mukohyama2010}.
They concluded that a spherically symmetric star should include a time-dependent region near the center.
Although we cannot deny that stars should be described by dynamical configurations,
the fact that we cannot find simple stationary and axisymmetric star solutions with the four-velocity generated by the Killing vectors will be an unattractive feature of this theory.

Greenwald, Papazoglou and Wang found static spherically symmetric solutions with a perfect fluid plus a heat flow along the radial direction 
and with a class of an anisotropic fluid
under the assumption that the spatial curvature is constant in a projectable theory without the detailed balance condition~\cite{greenwald_papazoglou_wang2010},
although it is doubtful that the constant-spatial-curvature solutions represent realistic stars.
This, however, implies that rotating star solutions 
with a perfect fluid plus a radial heat flow and with an anisotropic fluid 
can also exist.

We might get star solutions by introducing an exotic matter with a negative pressure,
but it seems that the physical justification to introduce it is difficult.

Our result does not imply the non-existence of rotation star solutions in this theory. 
However, it would be useful to investigate rotating-star solutions in this theory 
and then to compare the solutions with the corresponding ones in Einstein gravity for astrophysical tests of this theory.
Furthermore, although we do not disprove the existence of rotation star solutions with radial motion,
it is doubtful whether such star solutions describe realistic astrophysical stars.

Recently, the property of matter in the non-projectable version 
of the extended Ho\v{r}ava--Lifshitz gravity \cite{Blas_Pujolas_Sibiryakov_2010} 
at both classical and quantum levels has been investigated by Kimpton and Padilla~\cite{Kimpton:2013zb}. 
Although the gravity sector in Ho\v{r}ava--Lifshitz has been investigated eagerly, the matter sector has not, relatively. 
It is left as future work to answer the question of whether or not the no-go theorem applies at a quantum level.

\section*{Acknowledgements}
The authors would like to thank M. Saijo, U. Miyamoto, S. Kitamoto, N. Shibazaki,
T. Kuroki, S.~Mukohyama, K. Izumi, M. Nozawa, R. Nishikawa, M. Shimano, H. Nemoto, S. Kamata, S. Okuda and T. Wakabayashi for 
valuable comments and discussion.
Naoki Tsukamoto thanks the Yukawa Institute for Theoretical Physics at Kyoto University,
where this work was initiated during the YITP-W
-11-08 on ``Summer School on Astronomy and Astrophysics 2011''.
Tomohiro Harada was supported by the Grant-in-Aid for Scientific
Research Fund of the Ministry of Education, Culture, Sports, Science
and Technology, Japan [Young Scientists (B) 21740190]. 
\appendix

\section{Explicit Expression for Equation of Motion}

After a long straightforward calculation, we obtain the explicit expression for the equation of motion:
\begin{eqnarray}
\alpha \left [\frac{N}{2}K^{lm}K_{lm}g^{ij}-2NK^{im}K^j_m -\frac{1}{\sqrt{g}}(\sqrt{g}K^{ij}) 
-D_p(K^{ip}N^j)-D_p(K^{pj}N^i)+D_p(K^{ij}N^p)\right]\nonumber\\
-\alpha\lambda\left[\frac{N}{2}K^{2}g^{ij}-2NKK^{ij} -\frac{1}{\sqrt{g}}(\sqrt{g}Kg^{ij}) 
-D_p(Kg^{ip}N^j)-D_p(Kg^{jp}N^i)+D_p(KN^pg^{ij})\right]\nonumber\\
+\beta\left[-\frac{1}{2}NC^{kl}C_{kl}g^{ij}+2N C^{jl}C^i_l +2\varepsilon^{pkl}R^j_l D_k(NC^i_p) \right. 
-\varepsilon^{pki}D_m D^j D_k(NC^m_p) -\varepsilon^{pkl}D_l D^j D_k(NC^i_p)\nonumber\\
+\varepsilon^{pkj} D^lD_l D_k(NC^i_p)+\varepsilon^{pkl}g^{ij}D_m D_l D_k(NC^m_p)
-\varepsilon^{kil} D_p(NC^j_kR^p_l)-\varepsilon^{pkl} D_k(NC^j_pR^i_l)+\varepsilon^{pil} D_k(NC^k_pR^j_l) \left. \right]\nonumber\\
+\gamma \left[ \varepsilon^{pqk}D_p D^i(N D_qR^j_k+\frac{1}{2}R^j_kD_qN)
+ \varepsilon^{jqk}D_l D^i(N D_qR^l_k+\frac{1}{2}R^l_kD_qN)
-\varepsilon^{iqk}D^lD_l (N D_qR^j_k+\frac{1}{2}R^j_kD_qN) \right. \nonumber\\
\left. -\varepsilon^{pqk}g^{ij}D_p D_l(N D_qR^l_k+\frac{1}{2}R^l_kD_qN) 
+\varepsilon^{pqk}R^j_kD_q (N D_qR^i_p)+\varepsilon^{ikp}D_l (N R^l_pR^j_k) \right] \nonumber\\
+\zeta\left[\frac{1}{2}NR_{kl}R^{kl}g^{ij}-2NR^{il}R^j_l+2D_kD^j(NR^{ki}) 
-D^lD_l (NR^{ij})-g^{ij}D_{k}D_l(NR^{kl}) \right]\nonumber\\
+\eta\left[\frac{1}{2}NR^2g^{ij}-2NRR^{ij}+2D^iD^j(NR)-2g^{ij}D^lD_l (NR) \right] \nonumber\\
+\xi\left[\frac{1}{2}NRg^{ij}-NR^{ij}+D^jD^iN-g^{ij}D^lD_l N \right]
+\sigma N\frac{1}{2}g^{ij}+(i\leftrightarrow j)+\frac{2}{\sqrt{g}}\frac{\delta I_m}{\delta g_{ij}}=0 \nonumber\\
\end{eqnarray}
where $(i\leftrightarrow j)$ means the terms, $i$ and $j$, exchanged each other.

\section{Triad Components of Extrinsic Curvature Tensor}

In this theory, the triad components of the extrinsic curvature tensor also should be regular.
The Lie derivative of $g_{ij}$ along $N^{i}$ is: 
\begin{eqnarray}\label{eq:Lie-spatial-metric}
\mathcal{L}_\textrm{\boldmath $N$}g_{ij}
&=&D_{j}N_{i}+D_{i}N_{j} \nonumber\\
&=&g_{ik}N^{k}_{\;,j}+g_{jk}N^{k}_{\; ,i}+g_{ij,k}N^{k}
\end{eqnarray}
The extrinsic curvature tensor (\ref{eq:extrinsic-curvature}) and (\ref{eq:Lie-spatial-metric}) yield:
\begin{eqnarray}\label{eq:convective-time-derivative-of-spatial-metric}
\frac{dg_{ij}}{dt}-N^{k}_{\; ,i}g_{jk}-N^{k}_{\;,j}g_{ki}=2NK_{ij} 
\end{eqnarray}
where:
\begin{eqnarray}
\frac{d}{dt}\equiv \frac{\partial}{\partial t}-N^{i}\frac{\partial}{\partial x^{i}}
\end{eqnarray}
By projecting Equation (\ref{eq:convective-time-derivative-of-spatial-metric}) onto the triad (\ref{eq:triad-one})--(\ref{eq:triad-three}),
we obtain the following equations \cite{bardeen_piran1983}:
\begin{eqnarray}\label{eq:extrinsic_curvature_tensor_traid_11}
&&NK_{(1)(1)}=-N^{r}_{,r}+\frac{1}{A}\frac{dA}{dt}+\frac{2}{\psi}\frac{d\psi}{dt} \\
&&\frac{2NK_{(1)(2)}}{\sin\theta}=\frac{AB}{r}N^{r}_{\; ,X}-\frac{r}{AB\sin\theta}N^{\theta}_{,r} \\
&&\frac{2NK_{(1)(3)}}{\sin\theta}=-\frac{rB}{A}[N^{\phi}_{\; ,r}+\frac{\xi}{\sin\theta}N^{\theta}_{,r}] \\
&&NK_{(2)(2)}=\frac{1}{r}\frac{dr}{dt}+\frac{2}{\psi}\frac{d\psi}{dt}-\frac{1}{B}\frac{dB}{dt}-N^{\theta}_{\; ,\theta} \\
&&NK_{(3)(3)}=\frac{1}{r}\frac{dr}{dt}+\frac{2}{\psi}\frac{d\psi}{dt}+\frac{1}{B}\frac{dB}{dt}-\frac{\cos\theta}{\sin\theta}N^{\theta} \\ \label{eq:extrinsic_curvature_tensor_traid_23}
&&2NK_{(2)(3)}=B^{2}\frac{d\xi}{dt}+(1-X^{2})B^{2}(N^{\phi}_{\; ,X}+\frac{\xi}{\sin\theta}N^{\theta}_{,X})
\end{eqnarray}
where:
\begin{eqnarray}
X\equiv \cos\theta
\end{eqnarray}

\end{document}